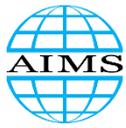

*Mini review*

# A literature review on COVID-19 disease diagnosis from respiratory sound data

**Kranthi Kumar Lella and Alphonse PJA\***

Department of Computer Applications, NIT Tiruchirappalli, Tamil Nadu, India-620015

**\* Correspondence:** Email: alphonse@nitt.edu; Tel: +917373774343.

**Abstract:** The World Health Organization (WHO) has announced a COVID-19 was a global pandemic in March 2020. It was initially started in china in the year 2019 December and affected an expanding number of nations in various countries in the last few months. In this particular situation, many techniques, methods, and AI-based classification algorithms are put in the spotlight in reacting to fight against it and reduce the rate of such a global health crisis. COVID-19's main signs are heavy temperature, different cough, cold, breathing shortness, and a combination of loss of sense of smell and chest tightness. The digital world is growing day by day; in this context digital stethoscope can read all of these symptoms and diagnose respiratory disease. In this study, we majorly focus on literature reviews of how SARS-CoV-2 is spreading and in-depth analysis of the diagnosis of COVID-19 disease from human respiratory sounds like cough, voice, and breath by analyzing the respiratory sound parameters. We hope this review will provide an initiative for the clinical scientists and researcher's community to initiate open access, scalable, and accessible work in the collective battle against COVID-19.

## 1. Introduction

As of 31st January 2021, the COVID-19 epidemic was declared a pandemic by the World Health Organization (WHO) in [1] the year 2020 of March 11, it claiming over 2,217,005 lives worldwide. The global situation as of 31st, January 2021, there have been 102,399,513 confirmed cases of COVID-19, which includes 2,217,005 deaths were reported to WHO is shown in Figure 1. Experts in microbiology believe that data collection is critical for isolating infected people, tracing connections, and slowing the spread of the virus. Although advancements in testing have made these methods more common in recent months, the need for affordable, fast, and scalable screening technology for COVID-19 is very much needed. The seriousness of COVID-19 disease is classified into three categories, namely extreme, middle/moderate, and mild. The problem of respiratory sound classification [2,3] and diagnosis of COVID-19 disease has received good attention from the clinical scientists and researchers community in the last year. In this situation, many AI-based models [4,5] entered into the real-world to solve such problems; and researchers have provided different machine learning, signal processing, and deep learning techniques to solve the real-world problem [6].

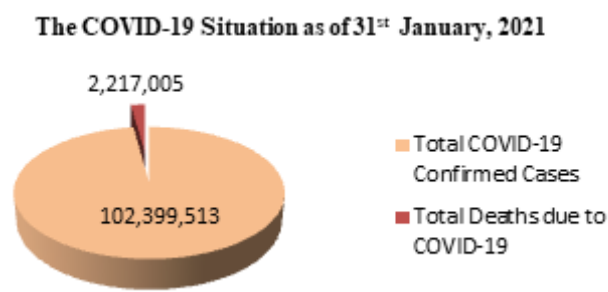

**Figure 1.** The global situation as of 31st January 2021 was reported to WHO.

Medical researchers and clinicians have used the audio sound produced by the human body, such as breath, respiratory sound, heart sound, swallow breathing, pulse sound, pulmonary sounds, etc., to diagnose and track human disease. Normally these symptoms were gathered by physical auscultation before recent patient visits. Researchers and scientists are just starting to use digital technologies to capture sounds from the human body by using digital stethoscopes and run automatic analysis on the data to detect the disease, for example, wheeze detection in asthma. The use of human speech/voice to aid early detection of several diseases (voice pathology detection, dry and wet cough, Parkinson's, depression, Alzheimer's, Migraines, etc.) [7] has also been piloted by researchers and scientists. Coronary artery disease speech frequency (hardening of the arteries that can impair the development of voice) and vocal tone, pitch, rhythm, rate, and volume correlated with an invisible disease such as posttraumatic stress disorder, traumatic brain injury, and psychiatric disorders.

The use of human-generated sounds as a diagnostic tool for different diseases presentments tremendous potential for early detection as well as an inexpensive solution that is embedded in commodity devices that could be rolled out to the people. It is even truer if these solutions unobtrusively cloud track individuals during their everyday lives. A recent study has started to examine how human respiratory sounds such as cough, voice recorded and breathing from hospital-tested COVID-19 positive devices vary from healthy people's sounds. A cough-based diagnosis of COVID-19 will also have to be taken into account with the respiratory and non-respiratory sounds data associated with all the conditions are mentioned [8]. Data review of a large crowdsourced respiratory speech/sounds dataset that has been obtained to accurately diagnose COVID positive cases was mentioned in [9]. It explains cough and respiration to understand how noticeable the sounds of COVID-19 are from those in asthma or protection of individuals. Speech recordings from hospital patients with COVID-19 are analyzed in [10] to categorize patients' health status automatically. The digital stethoscope, lung sounds data is being used in [11] as a diagnostic signal for COVID-19; in [12], the COVID-19 related cough detection analysis obtained by telephone is presented using a cohort of nearly 50 patients with COVID-19 versus other pathological coughs trained in a series of models.

*1.1. Spreading coronavirus (SARS-CoV-2) from COVID-19 biomedical waste*

Several pieces of researches have shown that the novel COVID-19 disease is spreading not only because of traditionally direct contact [13] but also by the COVID-19 biomedical wastage. Ilyas Sadia et al. proposed the methods and procedures used in COVID-19 hospitals to handle biomedical waste [14]. But after the pandemic of COVID-19, new techniques of biomedical waste have emerged, and this disinfection of COVID-19 waste is essential to manage the widespread of COVID-19. Ramtek Shobhana and LS Bharat have published a work [15] on COVID-19 disease outbreak implications for the biomedical waste industry in India to control the transmission of coronavirus. The possible consequences of the COVID-19 disease outbreak on healthcare waste management were addressed by the authors and emphasize fundamental focuses where optional working strategy or additional control steps might be necessary.

Arghya Das et al. published a letter [16] to the editor on problems in healthcare waste management for COVID-19 and suggested guidance for developing countries such as India. In addition to the guidelines to adapt current policies of medical waste management laws, such recommendations supported the use of doubly layered containers (using two or more bags), compulsory marking of containers and bags as COVID-19 garbage, periodic disinfection of specific containers, different record-keeping of waste created from isolation wards of COVID-19 hospitals. Filimonau Viachaslau published work on [17] the prospects for medical waste management in the healthcare sector of SARS-CoV-2 after the pandemic and discussed the problem of plastic waste, and the healthcare sector should develop in 'green' innovations. The authors suggested that potential solutions to better mitigate this waste in the hospitality field in a post-pandemic environment. The hospitality industry should be incorporated into alternative, short-term food supply systems and food channels to resolve the problem of food waste.

Kulakarni Bhargavi N and Anatharama V summarized [18] the municipality's work on biomedical waste management problems and risks in the COVID-19 disease outbreak situation. It addressed the global waste management context during the COVID-19 outbreak and explored different aspects of biomedical waste management. The discussion enables the determination of infectious disease parameters by solid waste management, the implications of existing municipal waste treatment and disposal schemes for medical waste surgeons. The authors proposed alternative methods for waste management treatment to dumping waste and guidelines for the potential scope of tasks to develop a proper waste management environment during and aftermath of the COVID-19 pandemic. During and after the SARS-CoV-2 disease outbreak, Sharma Hari Bhakta et al. presented a report [19] on possibilities, obstacles, and technologies for efficient and proper waste management, as well as authors addresses particular cases for plastic waste, pharmaceutical waste, and food waste management and also addresses the need for distributed robust supply chains to be installed during future epidemics to tackle such circumstances.

In the background of the SARS-CoV-2 pandemic, Ganguly Ram Kumar and Chakraborty SK introduced an integrated framework for the management of urban solid waste [20] and discussed the problems faced by the current waste management system to tackle the massive generation of waste. In the light of an emerging disease outbreak, the authors addressed all newly created challenges with the need to outline strategies that combine various conventional, innovative, and newly suggested waste management strategy to manage the emerging environmental disasters, particularly concerning the collection, processing, disposal and recycling of enormous amounts of municipal solid waste. In reference to the COVID-19 pandemic, Jean Philippe Adam et al. reported [21] the observations of the hospital pharmacy of the Centre Hospital University Montreal (CHUM). They addressed the seven main issues: internal communications, virtual organizations, risk management, time management, employee tension, pharmacist staff reorganization in pharmaceuticals, and reorganizing of workplaces. The biomedical waste management technique of COVID-19 is challenging to reduce health consequences and can be of great importance to developing new strategies for the potential control/prevention of the SARS-CoV-2 pandemic.

*1.2. COVID-19 sounds taxonomy*

The new Information and Communication Technologies (ICT) support fighting against COVID-19 in several directions, and it included attempts at analysis towards:
1. Classification of COVID-19 symptoms with abnormal respiratory patterns [2].
2. Diagnosis and treatment for COVID-19 by using ML and DL technique [6].
3. Identification of COVID-19 symptoms with cough data through mobile application [8].
4. Crowd-sourced respiratory sound data to the diagnosis of COVID-19 [9].
5. COVID-19 detection with cough using "COUGHVID" crowdsourcing dataset [22].
6. Correlation analysis of COVID-19 sounds with MFCC method [23].
7. Diagnosis of COVID-19 by analyzing pulmonary voice vocal fold oscillations [24].
8. AI for detection of COVID-19 with respiratory cough sound [25,26].
9. Biomarkers framework for Detection of COVID-19 with speech-production subsystems [27].
10. Speech analysis under COVID-19 with parameters [28].

The taxonomy of respiratory COVID-19 sounds from crowdsourced datasets is depicted in Figure 2, and we have collected datasets related to COVID-19 respiratory sounds (voice, cough, breath) from crowdsourcing datasets and data-driven techniques such as Artificial Intelligence Machine Learning and Deep Learning techniques to diagnose COVID19 disease. These techniques and methods help to detect the COVID-19 symptoms from the crowdsourced respiratory sound data. We can detect the COVID-19 positive case symptoms with cough sound, screening of a patient breath with speech results, and an Artificial Intelligence (AI) machine can sense COVID-19 symptoms from continuous speech and mental health situation.

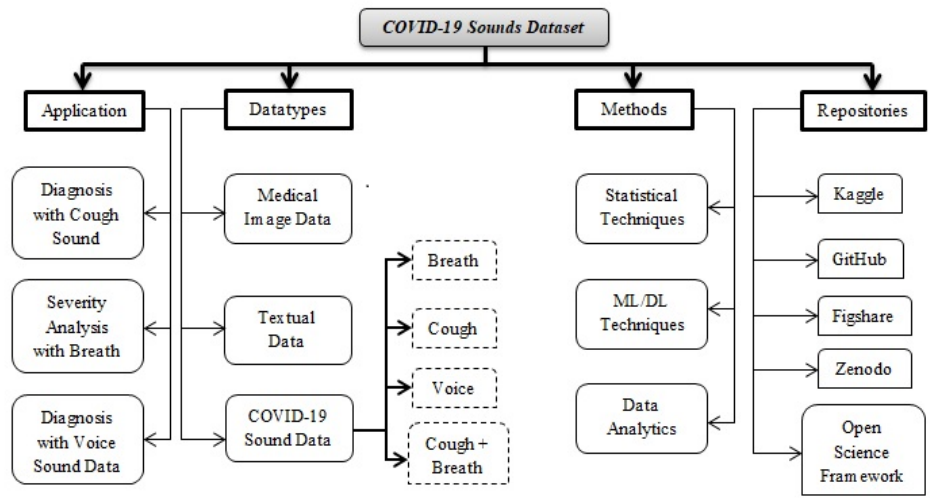

**Figure 2.** Taxonomy of respiratory COVID-19 sounds from crowdsourced datasets.

Artificial intelligence (AI) is one method that could well help sustain this endeavor to assess the researchers and scientists for prescreening of COVID-19. To the best of our knowledge, this one is the first review article on the subject of this study. In this paper, we briefly describe the COVID-19 data and COVID-19 sound analysis, and the general taxonomy of respiratory COVID-19 sounds from crowdsourced datasets is in the introduction section 1. Background works related to COVID-19 respiratory sounds data and the contribution of each author's work described in section 2. The results comparison analysis of each author and accuracy performance on different datasets is explained in section 3, and we have concluded our literature review analysis with the summary and future directions.

## 2. Literature review and background work

The utility of sound has long been recognized by researchers and scientists as a potential predictor of actions and health. For example, purpose-built external microphone recorders have been used in digital stethoscopes to detect sounds from the heart or lungs. These also involve highly trained clinicians to listen and interpret and are recently and quickly being replaced by various technologies such as a variety of imaging techniques (e.g., MRI, sonography) with which it is easier to examine and interpret. Recent trends in automated audio interpretation and modeling, however, have the potential to reverse this trend and gives sound as an alternative that is inexpensive and easily distributed. More recently, microphones have been exploited for sound processing on commodity devices such as smartphones and wearables.

In [2], Wang Yunlu et al. proposed a method to classify large-scale screening of people infected with COVID-19 differently; this work can be used to identify various breathing patterns and we can bring this tool for practical use in the real world. In this paper, first, a new and strong RS (Respiratory Simulation) Model is introduced to fill the gap between a huge amount of training data and inadequate actual data from the real-world to considering the characteristics of real respiratory signals. To identify six clinically important respiratory patterns, they first applied bidirectional neural networks like the GRU network attentional tool (BI_at_GRU) (Tachypnea,

Eupnea, Biots, Cheyne-Stokes, Bradypnea, and Central-Apnea). The research results show that six distinct respiratory trends with 94.5%, 94.4%, 95.1%, and 94.8% respectively, precision, accuracy, recall, and $F_1$ can be identified by the proposed model. In comparative studies, the acquired BI_at_GRU specific to the classification of respiratory patterns outperforms the existing state-of-the-art models. The proposed deep model and design concepts have enormous potential to be applied to large-scale applications such as sleeping situations, public environments, and the working environment.

In [3], a portable non-contact system was proposed by Jiang Zheng et al. to track the health status of individuals wearing masks by examining characteristics of the respiratory system. This device consists mainly of a thermal imaging camera with FLIR (Forward-looking infrared) and an Android device. Under realistic situations such as pre-screening in institutions and clinical centers, and this can help distinguish those possible COVID-19 patients. In this work, they performed health screening with the combination of thermal and RGB videos from DL architecture-based cameras. Firstly, they were used pulmonary data analysis techniques to recognize mask-wearing people; to obtain the health screening outcome, a BI_at_GRU function is applied on pulmonary disease results, and finally, they achieved 83.7% accuracy to classify the respiratory health conditions of a diseased patient.

In [8], Imran Ali et al. implemented an AI (Artificial Intelligence) based screening solution to detect COVID, transferable through a smart mobile phone application was suggested, developed, and finally tested. The mobile App called AI4COVID-19 records and sends to AI-based clouds running in the cloud triple 3-second cough sounds and comeback reaction within two minutes. Generally, cough is a basic indication of over 30 medical conditions associated with non-COVID-19. This makes it an incredibly difficult multidisciplinary issue is cough alone to diagnose COVID disease. By investigating morphological direction changes with dissimilarities from cough respiratory achieves an accuracy of 88.76%.

In [9], Brown Chloe et al. proposes an Android/iOS app to collect COVID-19 sounds data from crowdsourced sounds respiratory data of more than 200 positives for COVID-19 from more than 7k unique users; Brown Chloe et al. has taken many general parameters and three major set COVID-19 tasks based on breath and cough sound. Here parameters are, i) COVID-positive/non-COVID, ii) COVID-positive with cough/non-COVID with cough, iii) COVID positive with cough/non-COVID asthma cough; in task one achieved 80% of accuracy for 220 users with modality is cough + breath; in task two achieved 82% of accuracy for 29 users with modality is cough only; finally in task three achieved 80% of accuracy for 18 users with modality is breath. Recall function is slightly lower (72%) because of the not specialized net to detect every COVID-19 cough.

Hassan Abdelfatah et al. [10] implemented a system to diagnose COVID positive by using the RNN model; authors illustrated the major impact of RNN (Recurrent Neural Network) with the use of SSP (Speech Signal Processing) to detect the disease and specifically, this LSTM (Long Short-Term Memory) used to evaluate the acoustic characteristics of patients' cough, breathing, and voice, in the process of early screening and diagnosing the COVID-19 virus. Compared to both coughing and breathing sound recordings, the model findings indicate poor precision in the speech test.

In [22], Orlandic Lara et al. implemented the "COUGHVID" crowdsourced dataset for cough analysis in COVID-19 symptom; More than 20,000 crowdsourced cough recordings reflecting a broad range of topic gender, age, geographic locations, and COVID-19 status are given in the COUGHVID dataset. They have collected a series of 121 cough sounds and 94 no-cough sounds first-hand to train the classifier including voice, laughter, silence, and various background noises. They have taken self-reported status variables (25% of recording sounds with healthy values, 25% sound recordings with COVID values, 35% sound recordings with symptomatic value, and 15% sounds recordings with non-reported status; It ensured that all three reviewers labeled 15% of cough sounds) for the selection of the recordings to be labeled. The percentage of COVID positive, Symptoms of COVID, and healthy subjects were 7.5%, 15.5, and 77% from the subject of 65.5% males and 34.5% females respectively. It generated as audible dyspnea (93.0%), wheezing (90%), stridor (98.7%), choking (99.1%), or nasal congestion (99.2%) from 632 labeled COVID-19 cough records.

In [23], Mohamed Bader et al. proposed the significant model with the combination of Mel-Frequency Cepstral Coefficients (MFCCs) and SSP (Speech Signal Processing) to the extraction of samples from non-COVID and COVID, and it finds the person correlation from their relationship coefficients. These findings indicate high similarity between various breathing respiratory sounds and COVID cough sounds in MFCCs, although MFCC speech is more robust between non-COVID-19 samples and COVID-19 samples. Besides, these findings are provisional, and it is possible to remove the various patient voices with COVID-19 for future analysis. They have collected three female and four male voices from seven healthy patients, and two female, five male voices from 7 COVID-19 patients were obtained from their dataset. They were obtained COVID-19 infected patients' data from Zulekha hospital in Sharjah. The data is four times cough from each speaker, the voice of numbers counting from 1 to 10 of each speaker, and 4 to 5 times deep breath of each speaker. In addition, when recording their speech signals, the patients must sit with their heads straight in a comfortable way; three recordings for each speaker are acquired from smartphone devices in data collection, which can affect the quality of sound.

Mahmoud Al Ismail et al. [24] proposed a model with an analysis of vocal fold oscillation to detect COVID-19; because most symptomatic COVID-19 patients have mild to extreme impairment of respiratory functions, we hypothesize that through analyzing the movements of the vocal folds, COVID-19 signatures might be detectable. Our objective is to confirm this hypothesis and quantitatively characterize the changes observed to enable voice-based detection of COVID-19. We use a dynamic system model for vocal fold oscillation for this and use our recently developed ADLES algorithm to solve it to generate vocal fold oscillation patterns directly from recorded speech. Experimental findings on COVID-19 positive and negative subjects on a scientifically selected dataset show characteristic patterns of vocal fold oscillations associated with COVID-19. A data collection obtained under clinical supervision and curated by Merlin Inc., a private firm in Chile, was used for our research. The dataset contained recordings of 512 individuals who were tested for COVID-19, resulting in either positive or negative COVID-19 results. Among these, we only selected the recordings of those people who were reported within seven days after being medically examined. This criterion was met by only 19 citizens. Of these, there were ten females and nine males. COVID-19 was diagnosed in five women and four men, and the remainder tested negative. 91.20% is the efficiency of logistic regression on extended vowels and their combinations.

Chaudhari Gunavant et al. [25] their research show that crowdsourced cough audio samples collected worldwide on smartphones; various groups have gathered several COVID-19 cough recording datasets and used them to train machine learning models for COVID-19 detection. However, each of these models has been trained on data from a variety of formats and recording settings; collected additional counting and vocal recordings, authors exclusively collect cough recordings. Besides, these datasets come from different sources, such as collecting data from clinical environments, crowdsourcing, and public media interview extraction; It combined with COVID-19 status labels can be used to create an AI algorithm that correctly predicts COVID-19 infection with a 77.1% ROC-AUC (75.2–78.3%). In addition, without more training using the relevant samples, this AI algorithm can generalize to crowd-sourced samples from Latin America and clinical samples from South Asia.

Laguarta Jord et al. [26] proposed an AI (Artificial Intelligence) model from cough sound recordings to detect the COVID symptoms; this model allows a solution to prescreen COVID-19 sound samples country-wide with no cost. It achieves 97.1% of accuracy to predict the COVID positive symptom from cough sounds and 100% of accuracy to detect asymptomatic based on cough sounds of 5320 selected datasets.

Quatieri Thomas et al. [27] proposed a framework structure to identify COVID symptomatic condition with Signal Processing (SP) and Speech modeling techniques; this technique relies on the complexity of neuromata synchronization over speech/sound respiratory subsystem inside in the articulation, breathing, and phonation, driven by the existence of COVID symptom involving in upper inflammation versus lower respiratory inflammation tract. It is well-growing proof for pre-exposure of the COVID (pre-COVID) and post-COVID is given by the researcher analysis with voice meetings of 5 patients. This proposed method offers a possible capacity for flexible and continuous study to show the dynamics of patient activity in real-life settings for advanced warning and monitoring of COVID.

In [28], Jing Han et al. proposed a study of intelligent analysis on COVID-19 speech data by considering four parameters such as; i. Sleep Quality, ii. Severity, iii. Anxiety, iv. Fatigue. Jing Han et al. collected data from the "COVID-19 sounds app" has launched by scientists and researchers from Cambridge University and the "Corona voice detect App" has launched by researchers from Mellon University. After data processing, these people have obtained 378 total segments; from this preliminary study, they have taken 260 recordings for future analysis. These 256 sound pieces have been collected from 50 COVID-19 infected patients; for future study, poly impulses with such a sample rate of 0.016MHz are converted. They have considered two acoustic feature sets in this study, namely ComParE & eGeMAPS; both feature sets were achieved 69% accuracy.

Kota Venkata Sai Ritwik et al. [29] proposed and investigated the presence of signs in the speech data about the COVID-19 disease; it very closes approach for the speakers to accept. Each sentence of Mel filter bank features for each phoneme is represented as support vectors. A two-class classifier is used to acquire the features of COVID-19 speech from regular. The small size of video data was collected from YouTube and showed that 88.6% accuracy and 92.7% $F_1$-Score can be achieved by an SVM classifier on this dataset. Further analysis indicates that the two classes can be differentiated better than the others by certain telephone classes (stops, mid vowels, and nasals).

## 3. Results and discussion

Table 1 shown the different result analysis with different methods performed by different author's; Brown Chloe et al. performed PCA and SVM classifier on COVID-19 dataset to detect COVID-19 disease and achieved 80%, 82%, 80% of accuracies for 3 tasks (COVID Positive/Non-COVID, COVID Positive with Cough/Non-COVID with Cough, COVID Positive with cough/Non-COVID asthma cough) with three modalities (Cough + Breath, Cough, Breath). Jing Han et al. performed an SVM classifier to the diagnosis of COVID-19 disease on Corona voice detected data app and analyzed conditions of COVID patient concerning sleep, fatigue, and anxiety is recorded 57%, 50%, 50% accuracy. Orlandic Lara et al. performed PDS, down-sampling, low pass filter, and TPE methods to produce labeled data on the COUGHVID dataset; and finally, it produced 632 labeled COVID-19 cough records the accuracy of audible dyspnea (93%), wheezing (90.5%), stridor (98.7%), choking (99.1%), nasal congestion (99.2), and 86.2% of accuracy for labeled as mild.

**Table 1.** It shows the literature review analysis of different authors by considering methods, tasks, and modalities on different COVID-19 sounds datasets.

| Ref. | Author | Year | Dataset | Methods | Tasks | Modality/Parameters | Accuracy (%) |
|---|---|---|---|---|---|---|---|
| [2] | Wang Yublu et al. | Feb. 2020 | Fig-share dataset contains Shear force direction and vertical force direction. | BI-ATGRU | COVID-19 detection with two classes | Breathing Patterns | 94.50 |
| [3] | Zheng Jiang et al. | April 2020 | This data was collected from patients in the Department of Respiratory Diseases and the Department of Cardiology at Ruijin Hospital. | BI_at_GRU mechanism | COVID-19 Detection | Combination of breath and Thermal videos | 83.69 |
| [8] | Ali Imran et al. | June 2020 | ECS_50 Dataset and COVID samples collected from the App. | DTL-MC | COVID-19 cough samples | Cough | 92.85 |
| | | | | | | Waves from voice sound | 92.64 |
| | | | | | | Overall | 88.76 |
| [9] | Brown Chloe et al. | July 2020 | COVID-19 Sounds | PCA and SVM, VGGish | i. COVID-positive/non-COVID | 1. Cough +Breath | 80 |
| | | | | | ii. COVID-positive with cough/non-COVID with cough | 2. Cough | 82 |
| | | | | | iii. COVID positive with cough/non-COVID asthma cough | 3. Breath | 80 |

| Ref. | Author | Year | Dataset | Methods | Tasks | Modality/Parameters | Accuracy (%) |
|---|---|---|---|---|---|---|---|
| [10] | Hassan Abdelfatah et al. | Oct. 2020 | Own dataset is collected from 14 users (7 COVID patients and 7 are Non-COVID Patients) | LSTM | i. COVID-19 Cough Sound ii. Non-COVID-19 Cough Sound | Cough Sound | 97 |
|  |  |  |  |  |  | Breath Sound | 98.2 |
|  |  |  |  |  |  | Voice | 84.4 |
| [22] | Orlandic Lara et al. | Sept. 2020 | COUGHVID | PDS, down-sampling, low pass filter, TPE | Produced 632 labeled COVID-19 cough records | Audible dyspnea | 93 |
|  |  |  |  |  |  | wheezing | 90.5 |
|  |  |  |  |  |  | stridor | 98.7 |
|  |  |  |  |  |  | choking | 99.1 |
|  |  |  |  |  |  | nasal congestion | 99.2 |
|  |  |  |  |  | Labeled as Mild | ---- | 86.2 |
| [23] | Mohamed Bader et al. | Oct. 2020 | Own dataset is collected from 14 users (7 COVID patients and 7 are Non-COVID Patients) | MFCC | i. Non-COVID-19 Vs COVID-19 ii. COVID-19 Vs COVID-19 | Breath | 43 |
|  |  |  |  |  |  | Cough | 42 |
|  |  |  |  |  |  | Voice | 79 |
|  |  |  |  |  |  | Breath | 58 |
|  |  |  |  |  |  | Cough | 65 |
| [24] | Mohmoud AI Ismail et al. | Oct. 2020 | Dataset of COVID-19 positive and negative | LR, NL-SVM, DT, RF, AB | Classifiers are ROC-AUC | LR | 82.5 |
|  |  |  |  |  |  | NL-SVM | 78.9 |
|  |  |  |  |  |  | DT | 80.3 |
|  |  |  |  |  |  | RF | 79.4 |
|  |  |  |  |  |  | AB | 81.2 |
| [25] | Chaudhari Gunvant et al. | Nov. 2020 | Collected data from clinical environments, crowdsourcing, and public media interview extraction | MFCC + Spectrum +Feature Extraction | i. Negative cough ii. Positive cough | Coughvid | 77.1 |
|  |  |  |  |  |  | Virufy Crowdsourced | 72.1 |
|  |  |  |  |  |  | Clinical Dataset1 | 58.6 |
|  |  |  |  |  |  | Clinical Dataset2 | 71.8 |
| [26] | Laguarta Jord et al. | Oct. 2020 | MIT open voice model (5,320 subjects have recorded a healthy COVID-19 cough dataset.) | MFCC, ResNet50 (CNN) | i. COVID Positive ii. COVID Negative | Personal Assessment | 79.2 |
|  |  |  |  |  |  | Doctor Assessment | 96.7 |
|  |  |  |  |  |  | Official Test | 97.1 |

| Ref. | Author | Year | Dataset | Methods | Tasks | Modality/Parameters | Accuracy (%) |
|---|---|---|---|---|---|---|---|
| [27] | Quatieri Thomas et al. | May 2020 | YouTube, Instagram, Twitter outlets audio data for five subjects. | Speech production model | i. Pre_COVID  ii. Post_COVID | - | - |
| [28] | Jing Han et al. | May 2020 | COVID-19 Sounds Data set, Corona Voice Detect Data app. | SVM | Sleep | eGeMAPS | 57 |
| | | | | | | ComPARE | 39 |
| | | | | | Fatigue | eGeMAPS | 50 |
| | | | | | | ComPARE | 44 |
| | | | | | Anxiety | eGeMAPS | 50 |
| | | | | | | ComPARE | 52 |
| [29] | Kotra Venkata Sai Ritwik et al. | Nov. 2020 | Small Dataset Collected from YouTube Videos | SVM | Two Classes, i. COVID-19 Positive Speaker | SVM | 88.6 |
| | | | | | ii. COVID-19 Negative Speaker | SVM F Score | 92.7 |

*Note: BI_at_GRU-Bi-directional_and_Attentional_Gated_Recurrent_Unit Neural Network; DT: Decision tree; DTL_MC: Deep_Transfer_Learning based Multi-Class_classifier; NL_SVM: Non-Linear_Support_Vector_Machine; MFCC: Mel_Frequency_Cepstral_Coefficients; LR: Linear Regression; AB: AdaBoost; ROC_AUC: Receiver Operating Characteristics_ Area Under Curve; RF: Random Forest.

Wang Yunlu et al. have taken the fig_share dataset with two classes and performed BI-ATGRU; achieved 94.5% accuracy for breathing patterns. Imran Ali et al. applied DTL-MC based classifier on the ECS_50 dataset (for training), COVID samples collected from the data app to diagnose COVID-19 by investigating dissimilarities of morphological direction changes in the respiratory system with the modality of two cough classes and four sound wave classes achieved overall 88.76% of accuracy. Jiang Zheng et al. performed Bidirectional Gated Recurrent Unit with an attenuation mechanism on patient respiratory data at Ruijin Hospital, shanghai; it achieves an accuracy of 83.69% for breathing & thermal videos. Mohamed Bader et al. performed MFCC technique on 14 patients data (7 are COVID and 7 are Non-COVID) with two different classes (Non-COVID Vs. COVID, COVID Vs. COVID); and achieved accuracy for class 1 is breath 43%, cough 42%, voice 79% respectively class 2 achieves breath 58%, cough 65%, and voice 82% of accuracy. Jiang X et al. performed MFCC, Spectrum, and Feature Extraction methods on collected data from clinical environments crowed searching in public media; and achieved 72.1% (best) accuracy for negative cough and positive cough classes. Kotra Venkata Sai Ritwik et al. performed an SVM classifier on the YouTube Videos dataset to classify COVID positive patients with the non-COVID patients and achieved 88.66% accuracy.

Laguarta Jord et al. performed MFCC, ResNet50 classifiers to classify COVID positive and negative patients on MIT open voice data model; and achieved an accuracy of 79.2%, 96.7% 97.1%, from personal assessment, doctor assessment, and official test. Hassan Abdelfatah et al. applied RNN (Recurrent Neural Network) based LSTM (Long Short-Term Memory) on 14 patients' data (7-positive and 7-negative) to identify Non-COVID-19 cough sound with COVID-19 cough 151sound, and it achieves 84.4% of accuracy for LSTM. The result analysis with accuracy, datasets what we have collected, and techniques to detect COVID-19 is depicted in Table 1. The existing works are not enough for the diagnosis of COVID-19 disease using human respiratory sounds; researchers and scientists have to propose more innovative AI methods and techniques to improve the system performance to the diagnosis of COVID-19 from respiratory sounds data. This literature review may help the researchers and clinical scientists to move their research direction towards this area.

## 4. Conclusion

There is no one who has not felt its influence in some way as COVID-19 has filled across the world. Researchers, experts in the medical industry, and biomedical scientists are now increasingly working to find a suitable solution for the disease. Artificial intelligence (AI) is one method that could well help sustain this endeavor to assess the researchers and scientists. In this review article, we majorly focus on literature analysis for COVID-19 disease diagnosis through COVID-19 respiratory sounds data with various AI-based techniques, and we compared results and techniques developed by each author. These techniques help researchers and clinical scientists to move towards their research directions. We observed that it is critical to find the COVID-19 positive disease with respiratory sounds of humans. So, the AI-based methods are appropriate, reliable, and effective in diagnosis COVID-19 disease from respiratory sounds. Cooperation between AI-based research scientists, medical researchers, and government organizations is required to do this. Even more, research should also recognize empirical AI models on physical world data with real-world examples. This paper has taken initial measures to collect and illustrate the existing state-of-the-art, but it does not distinguish between in the wild' working examples as well as those performed under laboratory conditions. This paper defines an initial measurement to collect reviews to the diagnosis of COVID-19 from patient respiratory sounds data, and it drives researchers to moves towards these directions. In these ways, we study and critically analyze practices that are supported by open access and technology. We hope the review will provide an initiative for the clinical scientists and researcher's community to initiate open access, scalable, and accessible work in the collective battle against COVID-19.

Feature challenges and perspective: Future outcomes on classification and regression techniques to diagnosing COVID-19 diseases from respiratory sound data have been developed in the biomedical and engineering field. To address a wide variety of clinical and bioengineering problems, researchers have exploited AI-based techniques. We have seen many AI approaches for diagnosing COVID-19 disease with data on respiratory sounds, and we will use the CNN (Convolutional Neural Network) model on COVID-19 crowdsource sounds dataset [9] using pre-processed data with the Data De-noising Auto Encoder (DDAE) technique to improve the better performance to the diagnosis of COVID-19 disease in future work.

**Conflict of interest**

The authors declared no conflict of interest.

*AIMS Bioengineering*                                                                                                           Volume 8, Issue 2, 140–153.

**Author contributions**

Kranthi Kumar Lella: Original draft preparation, Formal Analysis, and Data Conceptualization. Alphonse PJA: Supervision, Reviewing, and Editing.